\documentstyle[prd,preprint,aps]{revtex}
\newcommand{\be}{\begin{equation}}
\newcommand{\ee}{\end{equation}}
\newcommand{\bea}{\begin{eqnarray}}
\newcommand{\eea}{\end{eqnarray}}

\begin{document}
%%%%%%%%%%%%%Title page%%%%%%%%%%%%%%%%%%%

\reversemarginpar
\tighten
\title{Horizon Dynamics of a  BTZ Black Hole}
\author{A.J.M. Medved}
\address{Department of Physics and Theoretical Physics Institute\\
University of Alberta \\
Edmonton, Canada T6G-2J1\\
e-mail: amedved@phys.ualberta.ca}
\maketitle
\begin{abstract}

It has been suggested in the literature that, given
a  black hole spacetime, a relativistic
membrane can  provide an effective description
of the horizon dynamics.
  In this paper, we explore such a framework
 in the context of a 2+1-dimensional BTZ
black hole. Following  this membrane prescription,
we are able
to translate the horizon dynamics (now  described by a string)
into the convenient form of a 1+1-dimensional
Klein-Gordon equation. We proceed to
quantize the solutions and construct a thermodynamic
partition function. Ultimately, we
are able to extract the  quantum-corrected entropy, 
which is shown to comply with
the BTZ form of the Bekenstein-Hawking area law.  We also
substantiate that the leading-order correction  is  proportional
to the logarithm of the area.

\end{abstract}

\section{Introduction}

Many explorations into quantum gravity have  
centered in the realm of black hole thermodynamics \cite{WALD}.
In this regard, a particularly important open question 
is the  origin of the Bekenstein-Hawking entropy 
\cite{BEK,HAW},\footnote{Here,
$A$ is the area of the black hole horizon (or the analogue
of area when the dimensionality of the spacetime differs
from four), $G$ is the  gravitational coupling constant,
and all other fundamental constants have been set to unity.}
\be
S_{BH}= {A\over 4G}.
\label{11}
\ee
It is commonly believed that a derivation of the Bekenstein-Hawking
area law from first principles will be a significant step 
towards realizing the fundamental theory of quantum gravity \cite{ROV}.
Meanwhile, although this law is well established at the level of
semiclassical thermodynamics, the statistical origin
of the entropy still remains as enigmatic as ever.
\par
We should point out that
there has been undeniable success in  calculating  $S_{BH}$
by way of  state-counting procedures \cite{SV}.
 Nevertheless,
it is not at all evident that the states being counted  have any
real physical significance \cite{CARX}. This dilemma may be viewed
as a manifestation of our ignorance of physics below
the Planck scale. That is to say, without the resolution
of ``subplanckian'' distances, it seems unlikely
that theorists  will be able to identify, never mind count, the
microscopic degrees of freedom that (presumably) underlie the 
black hole entropy. 
\par
In view of this ignorance, it is perhaps beneficial
to ``take a step back'' and see what we can learn about black holes
when the subplanckian degrees of freedom have been
{\it a priori} suppressed.  To this end, a  very elegant
framework has  been  proposed by Maggiore \cite{MAG2}.   
This proposal
will serve as the focal point of our current analysis,
so  let us proceed  with a pertinent discussion.
\par
We begin by considering a fiducial observer; that is, a
static observer who remains eternally outside of the
black hole. It immediately follows from the ``no-hair'' theorem of
black holes \cite{BALD} that, as far as this
observer is concerned, only the degrees of freedom
{\it outside} of the horizon are of relevance.  Therefore,
we should limit considerations to the region of
spacetime that spans from the horizon
surface ($r=r_+$) to spatial infinity ($r\rightarrow\infty$). 
In terms of path integral formalism \cite{GH}, this implies
that the relevant partition function is expressible as follows:
\be
{\cal Z}=\int_{M_{ext}}{\cal D}[g_{\mu\nu}]e^{iI_{g}[g_{\mu\nu}]}.
\label{12}
\ee
Here,  $M_{ext}$ denotes the {\it exterior} manifold,
$g_{\mu\nu}$ is the metric for the background
(black hole) spacetime,
$I_g$ is the appropriate gravitational action,
 and ${\cal D}$ indicates
a suitable measure. 
\par
An immediate problem with the above picture is our  ignorance
with regard to the position of the horizon. (This is, of course,
essentially the same ignorance that has been alluded to above.)
At the classical level, we can, given $g_{\mu\nu}=g_{\mu\nu}^{cl}$, 
pinpoint the 
horizon precisely;
however,  at the quantum level, the metric is fluctuating and,
therefore, so is the position of the horizon. Let us assume
that the fluctuations  have a maximal spatial extent of
$\epsilon$, which is presumably on the order of a few
Planck lengths.  It is then natural 
to separate the exterior spacetime into a pair of
submanifolds that are defined by $r_+ < r < r_+ +\epsilon$
and $r> r_+ + \epsilon$.\footnote{Such a near-horizon cutoff in a 
black hole spacetime
is philosophically similar to the brick-wall model
proposed by 't Hooft \cite{BW}.} In the spirit of Wilson's renormalization
group \cite{WIL}, one can then integrate out the degrees of freedom
in the near-horizon shell ($M_\epsilon $) and employ the classical metric
on the outside. Following this  prescription,   
we can schematically re-express the partition function (\ref{12})
as follows:
\be
{\cal Z}= \left[ e^{iI_g[g_{\mu\nu}]}
\int_{M_{\epsilon}}{\cal D}[\xi]e^{iI_{\epsilon}[\xi]}
\right]_{g_{\mu\nu}=g_{\mu\nu}^{cl}}.
\label{13}
\ee
Here, $\xi$ collectively represents any physically relevant variables
that remain after integrating  the ``fast variables''
out of $M_{\epsilon}$, and $I_{\epsilon}[\xi]$ represents whatever 
effective action
has been induced by this coarse-graining process. 
\par
A cautionary comment is in order. It is implicit
in this procedure - for  which only a small fraction of
the spacetime is subject  to  quantization  - that the degrees
of freedom of a black hole spacetime are mostly localized within
a small region near the horizon.  Although this extreme degree of localization
may be intuitively unsettling,
 just such a notion  has, in fact, been frequently  advocated 
in the literature
({\it e.g.}, \cite{BW,BEKY}). Heuristically speaking,
this localization follows from the
immense gravitational blue-shifting of any  energy  
in the near-horizon vicinity.  
\par
In practice, it would be extremely difficult to
obtain an explicit  formulation of the effective action,
$I_{\epsilon}[\xi]$. Nevertheless, Maggiore has argued on
the grounds of invariance principles that, at least  to the lowest order,
the  renormalization group procedure
should induce the action of a relativistic bosonic membrane \cite{MAG2}.
That is, 
\be
I_{\epsilon}=- {\cal T}\int d^n\xi\sqrt{-h},
\label{14}
\ee
where  $n+1$ 
is the dimensionality of the black hole spacetime,
the $\xi$-variables now parametrize the $n$-dimensional world-volume  of
the membrane,  $h$ is the determinant of a suitably defined induced
metric, and ${\cal T}$ is the  tension of the membrane.
\par
The above formalism suggests an intriguing picture:
the  dynamics of the black hole
horizon can  effectively be described by 
a membrane whose  equilibrium position is at a distance 
 $\epsilon$ from the horizon.
This membrane position can, in fact, 
be identified with the so-called ``stretched horizon'' (see below)
of the black hole. Meanwhile, our ignorance of 
subplanckian physics is now encapsulated in the arbitrariness of 
the parameter ${\cal T}$;
that is,  the membrane tension.
Presumably, this lost information can be retrieved from a more
fundamental theory, but this is not necessary for semiclassical
considerations.
\par
It is worth noting that the above framework also follows
intuitively from the viewpoint of the membrane paradigm \cite{MP};
which stresses that, for a fiducial observer, the
black hole horizon behaves  as if it  were a real membrane
that is endowed with physical properties. (A useful definition
of these properties necessitates that the membrane
is moved out a small distance, which effectively describes
the location of the stretched horizon.)
On the other hand, a free-falling observer would see no membrane at all;
however,
  this apparent paradox has been
nicely resolved by the principle of black hole {\it complementarity} 
\cite{SUS2}. 
\par
That this membrane picture leads to a self-consistent
description of horizon dynamics was amply demonstrated
through the cited work of Maggiore \cite{MAG2}. 
In a  related study \cite{LOU}, Lousto applied
the membrane description in a novel way and demonstrated
that, for a ``conventional'' 
3+1-dimensional theory, the membrane fluctuations 
could be described by a
2+1-dimensional Klein-Gordon equation. This was followed by
a procedure of quantization and then a
thermodynamic analysis. Most notably, the leading-order entropy was found
to comply with the Bekenstein-Hawking area law. Higher-order
corrections were also considered.
\par
Lousto's verification of the area law can be viewed as highly non-trivial,
inasmuch as a  Klein-Gordon description of the horizon fluctuations could
not have been {\it a priori} anticipated. It should, therefore, be of
considerable interest to see if the basic outcomes persist for
more exotic black hole scenarios.  The purpose of the current paper
is to consider just such a scenario; in particular, the
BTZ model \cite{BTZ}, which describes   solutions
of 2+1-dimensional anti-de Sitter gravity that have all the properties
of black holes. Our choice is motivated, in part, by a subsequent
paper by Maggiore which  demonstrated that the general philosophy
can indeed be translated into a BTZ context \cite{MAG1}. 
(Note, however, that the
relativistic membrane is, in this case, a string.)
Furthermore, the BTZ black hole, although essentially a toy model,
has generated substantial interest in various aspects 
 of gravitational theory. For instance,
the BTZ solution is dually related to certain stringy black holes 
\cite{X1,X2},  
has played a featured role in  microscopic entropy calculations \cite{X3,X4},
and has served as a useful ``laboratory'' for studying one-loop 
thermodynamics \cite{X5,X6,X7,X8,X9,X10}.
\par
To further motivate our choice to study, in particular,
 the BTZ black hole, let us take note
of a relevant paper by Horowitz and Welch \cite{HW}. These
authors made the important observation that
the BTZ black hole is essentially equivalent, under an appropriate
duality, to  a three-dimensional black string solution.\footnote{It
should be stressed that no
such duality is apparent in the case of four dimensions of
spacetime. Hence, the current considerations are
indigenous to theories of gravity that can be cast, at least locally,
into a three-dimensional framework.}
(Especially pertinent  to this observation: 
 the  2+1-dimensional anti-de Sitter metric is
 the natural choice for formulating
 the SL(2,R) projection of  the
  Weiss-Zumino-Witten  model. Significantly, this {\it WZW} model
 uses a conformal field theory  to describe  string propagation.)
It is quite feasible that the string description of
 the current paper is some sort of semiclassical manifestation
of  the string in the WZW model. If this  relationship could  be rigorously
established, 
it would provide an intriguing physical motivation for
the Horowitz-Welch duality. We, perhaps boldy, suggest that the 
current treatment
can be viewed as a modest step
in this direction. 
(Note that other string theoretical descriptions
have been advocated for the BTZ black hole \cite{X4}, and it
remains an open question as to how any of the various interpretations
might be related.)
\par
The rest of the paper is organized as follows. 
In Section 2, we  consider  the action of a relativistic
string  embedded in the background of 
a BTZ black hole spacetime.  Keeping in mind that the
string serves as an effective description of horizon
dynamics, we are able to express the first-order
field equations in the form of a 1+1-dimensional
Klein-Gordon equation. In Section 3, we quantize 
the relevant solutions, which can be
identified with fluctuations in the string's radial position,
and obtain a discrete energy spectrum. We then go
on to construct a thermodynamic partition function,
from which the free energy, internal energy and entropy
are extracted. The resulting expression for the entropy is
discussed  in  detail. Section 4 ends
with a brief summary.

\section{Effective Action and Field Equations}

On the basis of our preceding discussion (also see \cite{MAG2,MAG1}),
we will  assume that the horizon dynamics of a BTZ black hole
can be effectively described by the action of a relativistic
(bosonic) string. That is,
\be
I=- {\cal T}\int d^2\xi\sqrt{-h}.
\label{21}
\ee
Here, $\xi^{i}=\left\{\tau,\sigma\right\}$ are the (1+1-dimensional)
world-volume coordinates and $h$ is the determinant of the
following induced metric:
\be
h_{ij}=g_{\mu\nu}{\partial X^{\mu}\over\partial\xi_i}
{\partial  X^{\nu}\over\partial\xi_j},
\label{22}
\ee
where $X^{\mu}=X^{\mu}(\tau,\sigma)$ describes the embedding
of the string in a 2+1-dimensional spacetime and
$g_{\mu\nu}$ is the target-space metric.
\par
The above {\it Nambu-Goto} action \cite{STRI} is known to
be equivalent to
\be
I=- {{\cal T}\over 2}\int d^2\xi\sqrt{-h} h^{ij}g_{\mu\nu}
\partial_iX^{\mu}
\partial_jX^{\nu},
\label{BLIP}
\ee
where $\partial_i=\partial/\partial\xi^i$.
Varying this form with respect to $X^{\mu}$,
we  obtain the following field equation:
\be
\partial_{i}\left[\sqrt{-h}h^{ij}g_{\mu\nu}
\partial_j X^{\nu}\right]-{1\over 2}\sqrt{-h}
h^{ij}\partial_iX^{\nu}
\partial_j X^{\rho}\partial_{\mu} g_{\nu\rho}
=0,
\label{23}
\ee
where $\partial_{\mu}=\partial/\partial X^{\mu}$.
\par
For  a target-space metric, we now specialize to the
curved background of a static BTZ black hole, and so  \cite{BTZ}
\be
ds^2_g=-U(r)dt^2+ {1\over U(r)}dr^2 +r^2d\phi^2,
\label{24}
\ee
such that
\be
U(r)={r^2\over l^2}-8GM={r^2\over l^2}-{r_{+}^2\over l^2}.
\label{25}
\ee
Here,  $G$ is the 2+1-dimensional Newton constant ({\it i.e.}, $G\sim
l_p$, a  Planck length), $M$
is the ADM black hole mass, $l$ is the curvature
radius ({\it i.e.}, $\Lambda=-l^{-2}$ is the cosmological
constant), and $r_+=\sqrt{8GMl^2}$ is the  radius of
the black hole horizon.
Note that the coordinate $\phi$ is identified 
with a period of $2\pi$. Also note that we assume a
semiclassical regime; meaning $M$ is large enough
so that  $r_+ >> l_p$.
\par
Next, let us  utilize the gauge symmetry of the system
and fix the coordinates as appropriate for
a fiducial (static, external) observer.
This choice immediately implies that $t$, $r$ and 
$\phi$ can be identified with $X^0$, $X^1$ and
$X^2$ (respectively). Moreover, the static nature 
and axial symmetry of the spacetime  naturally
leads to the following gauge-fixing conditions:
\be
t=X^{0}(\tau,\sigma)=\tau,
\label{29}
\ee
\be
\phi=X^{2}(\tau,\sigma)=\sigma.
\label{210}
\ee
Thus, all of the dynamics of the system are contained
within the yet-to-be-determined radial
function, $r=X^{1}(\tau,\sigma)$.
\par
As discussed in Section 1, this effective description
follows from the premise of  a fluctuating horizon with quantum fluctuations
on the order of a few Planck lengths. Moreover, the string
should maintain an equilibrium position at the order of unity
 (in Planck units) from the actual horizon,
given that short-distance effects have already been
accounted for via an implied coarse-graining procedure.
It is, therefore, appropriate
to write
\bea
r=X^{1}(\tau,\sigma)  &=& r_+ + \epsilon +\delta r(\tau,\sigma)
\nonumber \\
&=& r_e  +\delta r(\tau,\sigma),
\label{211}
\eea
where $r_e$ is the equilibrium position of
the string, while $\epsilon$ (a constant ``cutoff'' length)
and  $\delta r(\tau,\sigma)$ (a quantum fluctuation)
are both on the order of a few Planck lengths.
Alternatively, under our semiclassical assumption, 
both $\epsilon$ and  $\delta r <<r_+$.
\par
Incorporating the above formalism into Eq.(\ref{22})
for the induced metric, we find (up to the first
order in $\delta r$)
\be
ds_h^2=-\left[U(r_e)+U^{\prime}(r_e)\delta r\right]d\tau^2
+\left[r_e^2+2r_e\delta r\right]d\sigma^2,
\label{212}
\ee
where a prime denotes differentiation with respect to
$r$.
\par
The above form of the induced metric enables an
explicit evaluation of the  field equation (\ref{23}).
Doing so, we obtain for
the $r$, $t$ and $\phi$ components 
(respectively)
\bea
-{1\over U(r_e)}\partial^2_{\tau}(\delta r)+{1\over r_e^2}
\partial^2_{\sigma}(\delta r)
-\left[{U^{\prime}(r_e)\over r_e} -{U(r_e)\over r_e^2}
+{1\over 2}U^{\prime\prime}(r_e)\right]\delta r
\nonumber \\
 ={U(r_e)\over r_e}+{U^{\prime}(r_e)\over 2} 
+{\cal O}\left[(\delta r)^2 \right] ,
\label{213}
\eea
\be
\left[{U(r_e)\over r_e}
+{U^{\prime}(r_e)\over 2}\right]\partial_{\tau}(\delta r)
+{\cal O}\left[(\delta r)^2 \right]=0,
\label{214}
\ee
\be
\left[{U(r_e)\over r_e}
+{U^{\prime}(r_e)\over 2}\right]\partial_{\sigma}(\delta r)
+{\cal O}\left[(\delta r)^2 \right]=0.
\label{215}
\ee
\par
From the last pair of equations,
it is quite evident that, for a non-trivial solution of $\delta r$, 
the quantity inside of the square brackets
(in either equation) must vanish. Imposing
this constraint on the remaining field equation (\ref{213}),
we are left with 
\be
-\left[{1\over U(r_e)}\partial^2_{\tau}-{1\over r_e^2}
\partial^2_{\sigma} +\mu^2\right](\delta r)=0,
\label{217}
\ee
where:
\be
\mu^2={U^{\prime}(r_e)\over r_e} -{U(r_e)\over r_e^2}
+{1\over 2}U^{\prime\prime}(r_e).
\label{218}
\ee
\par
It is interesting that the above (\ref{217}) is simply a two-dimensional
Klein-Gordon equation, with the background metric
corresponding to the classical limit of the
induced metric; {\it cf}, Eq.(\ref{212}).
In this way, we can identify $\mu^2$ with the effective
mass (squared) of the first-order fluctuations,
$\delta r$.
\par
 For future reference, note that
\be
\mu^2={3\over l^2}+ {\cal O}[\epsilon],
\label{219}
\ee
where we have applied Eqs.(\ref{25}) and (\ref{211}).
One can imagine  generalizations of the
2+1-dimensional solution used here (for instance, a BTZ black
hole with charge \cite{CLE}).
However, the precise form of the effective mass is irrelevant
to later arguments, provided that $\mu^2$ remains
well-defined in the limiting cases of interest; namely, $\epsilon
\rightarrow 0$ and $r_+\rightarrow \infty$.

\section{Quantization and Thermodynamics}

To proceed, it is, of course, necessary to solve
the above Klein-Gordon equation  (\ref{217}).
For this purpose, let us first decompose the
fluctuation field as follows:
\be
\delta r(\tau,\sigma)=\sum_{m}e^{im\sigma}R_m(\tau),
\label{31}
\ee
where the periodicity of $\phi=\sigma$ imposes
that $m$  can only take on integral values.
The above form allows us to separate
the variables in Eq.(\ref{217})  and eventually obtain
\be
{\ddot R}_m(\tau)+\omega_{m}^2R_m(\tau)=0,
\label{32}
\ee
where
\be
\omega_m^2=U(r_e)\left[\mu^2+{m^2\over r_e^2}\right]
\label{33}
\ee
and a dot denotes differentiation with respect to
$\tau$.
\par
Eq.(\ref{32}) can readily be identified with the
equation of motion for a harmonic oscillator
at frequency $\omega_m$. Hence, the  quantization
of the modes, $R_m$,  yields the following
energy spectrum:
\be
E_{nm}=(n+{1\over 2})\omega_m, \quad\quad\quad n=0,1,2,.... 
\label{34}
\ee
It should be kept in mind that $\omega_m\sim \sqrt{U(r_e)}$,
which is just the Tolman red-shift factor
at  the stretched horizon of
the BTZ black hole. 
As $r_e$ approaches the true black hole horizon ($r_+$),
 this red shift goes to zero and Eq.(\ref{34})
should then be regarded as an energy continuum.  Hence,
it is really our ignorance of physics below the Planck scale
that necessitates a non-vanishing cutoff and, therefore,
 induces the discrete spacing between the energy levels.
To take it a step further, if the spacetime
is truly quantized below the Planck level, then Eq.(\ref{34})
can also be viewed as a manifestation of this effect.
A further point of interest is that the above energy levels
can be interpreted as a discrete spectrum for the mass
of a BTZ black hole.  Significantly, this complies with Bekenstein's 
 notion of black hole spectroscopy \cite{BEKS}.
\par
Given the above outcomes, it is natural to construct
a thermodynamic partition function
in the following manner:
\be
{\cal Z}=\prod_{m=-\infty}^{+\infty}\sum_{n=0}^{\infty} 
e^{-\beta(n+{1\over 2})\omega_m},
\label{35}
\ee
where $\beta$ is the inverse of the equilibrium temperature
(discussed below). Identifying the sum over $n$ as a geometric
series, we have
\be
{\cal Z}=\prod_{m=-\infty}^{+\infty}
{e^{-\beta\omega_m/2}\over 1-e^{-\beta\omega_m}}.
\label{36}
\ee
Alternatively, one can re-express this result in terms
of the (Helmholtz) free energy:
\be
F=-{1\over\beta}\ln{\cal Z}=
-{1\over\beta}\sum_{m=-\infty}^{+\infty}
\ln\left[{e^{-\beta\omega_m/2}\over 1-e^{-\beta\omega_m}}\right].
\label{37}
\ee
\par
The standard formula for the internal energy of a thermodynamic
system gives us
\be
{\cal E}={\partial(\beta F)\over\partial \beta}
= \sum_{m=-\infty}^{+\infty}\left[{\omega_m\over
e^{\beta\omega}-1}+{\omega_m\over 2}\right].
\label{38}
\ee
The first term (on the right-hand side) is the anticipated
Planckian or thermal spectrum, whereas the second
term is a divergent contribution that would likely
be removed upon a suitable process of renormalization.
\par
The associated entropy can also be obtained via
a standard relation, for which we find
\be
S=-\beta(F-{\cal E})=
\sum_{m=-\infty}^{+\infty}\left[{\beta\omega_m\over
e^{\beta\omega}-1}-\ln\left(1-e^{-\beta\omega_m}\right)\right].
\label{39}
\ee
Considering that the string lives in the vicinity of  the black hole
horizon, we expect  the spacing between adjacent energy levels
to be correspondingly small; {\it cf}, Eq.(\ref{34}) and
the subsequent discussion.
 It thus follows that the above summation 
 can be accurately evaluated as an
integral over $m$. That is,
\be
S={2r_e\over\beta\sqrt{U(r_e)}}\int^{\infty}
_{\beta\omega_0}dx\left[{x\over e^x-1}-\ln\left(1-e^{-x}\right)
\right],
\label{310}
\ee
where the integration variable has been changed for
convenience  and $\omega_0=\omega_{m=0}$.
\par
So far, the equilibrium value of the temperature,
$\beta^{-1}$, has been left unspecified. However,
it seems realistic that this value should be
closely related to the Hawking temperature of the BTZ
black hole: $T_{BTZ}=r_+/2\pi l^2$ \cite{BTZ}.
In fact, one would most naturally expect 
that $\beta^{-1}=T_{BTZ}+{\cal O}[\epsilon]$,
and we will assume that this is correct.
\par
Applying the above, and also
recalling that $\omega_0=\mu\sqrt{U(r_e)}$,
$\mu\sim l^{-1}$ and $U(r_e)\sim \epsilon r_+/ l^2$
({\it cf},  Eqs.(\ref{33},\ref{219},\ref{25},\ref{211})),
we have
\be
\beta\omega_0\sim\sqrt{{\epsilon\over r_+}}<<1.
\label{3105}
\ee
In light of this deduction, the above integral (\ref{310})
can readily be evaluated to yield
\be
S={4\zeta(2)r_e\over\beta\sqrt{U(r_e)}}
+\mu r_e\ln[\mu^2\beta^2U(r_e)]+{\cal O}[\beta\omega_0],
\label{311}
\ee
where we have discarded an irrelevant constant term.
\par
Let us first focus on the leading-order term, which 
will be denoted by $S_1$. Applying the formalism
of the last two paragraphs, we find that
\be
S_1 \approx {\eta r_+^{3/2}\over l \epsilon^{1/2}},
\label{312}
\ee
where $\eta$ represents a numerical factor of
order unity. 
\par
To make sense of this result, it is necessary that the
cutoff parameter, $\epsilon$, be re-expressed in terms
of an invariant, proper distance; say $y$. More
specifically,
\be
y=\int^{r_+ +\epsilon}_{r_+}{dr\over \sqrt{U(r)}}=
l\sqrt{2\epsilon\over r_+}+{\cal O}[\epsilon],
\label{313}
\ee
and so
\be
S_1 \approx \eta{r_+\over y}.
\label{314}
\ee
By hypothesis, we have  $y\sim l_p\sim G$,
so that $S_1$ is in agreement (up to a numerical factor
of ${\cal O}[1]$) with the Bekenstein-Hawking area law
of a BTZ black hole \cite{BTZ},
\be
S_{BH}={A_+\over 4 G},
\label{315}
\ee
where $A_+=2\pi r_+$ is the ``area'' of
the horizon in 2+1 dimensions.
\par
In view of
the factorization of classical and quantum path integrals 
({\it cf},  Eq.(\ref{13})),
one would actually expect  the {\it total} black hole
entropy to be given by a sum:   the tree-level
area law (\ref{315}) plus the entropy of the quantum fluctuations
(\ref{311}). This means that, to the leading quantum order,
we can write
\be
S_{tot}=S_{BH}+S_{1}= {A_+\over 4}\left[{1\over G}+{\eta\over y}\right].
\label{3155}
\ee
 It may appear, at a first glance,
somewhat problematic that $y$ can, {\it in principle},
be extrapolated to an infinitesimally small distance
({\it i.e.}, $y<<l_p\sim G$).
For such an extrapolation, the (total) black hole entropy
would  apparently diverge; however, even in this event, 
the precise Bekenstein-Hawking formula can still
effectively be preserved. 
This observation follows by virtue of the inverse gravitational
coupling ($G^{-1}$) always being uncertain 
up to a potentially infinite renormalization \cite{SUSU,BS}.
What is important, from our current perspective,
is that $S_1$ does indeed comply with the area law,
so that the leading-order effects can always be renormalized
away. 
\par
Incidentally, it can be (and has been \cite{MI}) argued
that, for a calculation of this nature, the  leading-order
quantum term should be viewed as the principal
source of black hole entropy rather than a ``supplement''
(as implied by Eq.(\ref{3155})). However, thanks
to the renormalizability of $G$, these two viewpoints
are operationally indistinguishable  at the
order of the area law.
\par
Let us now cast our attention on the first-order correction
to the area law, which  will be denoted by $S_2$.
On the basis of very general arguments  (with origins in either
state-counting 
\cite{CAR} or thermodynamic principles \cite{DAS}), 
  the leading-order correction
is expected to be directly proportional to the logarithm of
the horizon area. Moreover,
for a BTZ black hole in particular, this logarithmic
correction appears to have a prefactor of -3/2 \cite{CAR,DAS}.
\par
From an inspection of the second term in Eq.(\ref{311}),
it is clear that the inverted argument of the logarithm
is indeed  proportional to $r_+ \sim A_+$. More explicitly,
we find (up to irrelevant constants, higher-order corrections
and a $\ln\epsilon$ term which will be commented on below)
that 
\be
S_2\approx -\mu r_e\ln(A_+).
\label{316}
\ee
Substituting Eq.(\ref{219}), we then obtain the following:
\be
S_2 \approx \left[-{\sqrt{3}\over l}r_+ + {\cal O}(\epsilon)\right]\ln(A_+).
\label{317}
\ee
Here, we find that the prefactor coincides with the
prescribed value of -3/2 \cite{CAR,DAS} only for the very special
instance of $r_+ =\sqrt{3} l/2$.
We view this discrepancy  as support for the notion that
$S_1$ is a supplementary rather than principal source of
the black hole entropy (see the prior discussion).
That is to say, if $S_{BH}$ and $S_{1}$ are fundamentally
distinct quantities, then one would not {\it a priori} expect their
leading-order corrections to be in precise agreement.
\par
Recall  the importance, from a renormalization perspective,
that  $S_1$  be in compliance with the area law. It
is similarly important
 that the logarithmic prefactor in $S_2$ 
remains finite  as $\epsilon\rightarrow 0$.
The reason being that it is not at all clear that
a term  $\propto \ln(A_+)$ can in any way be renormalized.
In this regard, our above finding 
is quite reassuring. On the other hand, $S_2$ also
gives rise to a term $\propto \ln(\epsilon)$, which
is dangerously divergent. However,
it it expected that such a term can indeed be renormalized away
\cite{BS,WIN} and is, therefore, of no physical consequence.

\section{Conclusion}

In summary, we have been considering  Maggiore's ``membrane model''
\cite{MAG2,MAG1} in the context of a 2+1-dimensional
BTZ black hole \cite{BTZ}. The central idea is 
that a relativistic membrane (or string for
this BTZ scenario)
can effectively 
 describe
the horizon dynamics of a black hole. This, in turn, suggests
that the elusive quantum degrees of freedom (in a black hole spacetime)
can be identified with the  fluctuations of
a suitably defined  membrane or string. In the current study, we have
found that these fluctuations conform to a two-dimensional
Klein-Gordon equation. Moreover,  we have shown
that the associated solutions can be readily quantized, thus leading to
a discrete  spectrum of energies.  This formalism  was then used
to construct a thermodynamic partition function, from
which the ``quantum'' black hole entropy
could ultimately  be extracted.
At the leading order, we substantiated the Bekenstein-Hawking
area law (for a BTZ black hole \cite{BTZ}), which
indicates that the horizon dynamics effectively translate
into a renormalization of the gravitational coupling \cite{SUSU}.
We also verified  a next-to-leading-order
correction that 
is directly proportional to the logarithm of the horizon area.
Although the logarithmic prefactor did not generally agree
with some prior calculations \cite{CAR,DAS}, we have argued
that our result is still consistent with any {\it a priori}
expectations.
\par
It is worth re-emphasizing that the positive results of
this analysis  are highly non-trivial; insofar as
our calculation of the black hole entropy followed from
the analysis of a two-dimensional Klein-Gordon equation. 
Let us also remind the reader that similar outcomes
were obtained by Lousto \cite{LOU} in a 3+1-dimensional
context. By generalizing this prior treatment, we have provided further
support for the membrane interpretation of a black
hole horizon \cite{MP}. Significantly, this membrane
paradigm  has already served as an antecedent
for black hole complementarity \cite{SUS2} and the 
holographic principle \cite{THO}; both of which
play a pivotal role in our current  understanding
of quantum gravity.

\section{Acknowledgments}

The author would like to thank V.P. Frolov for
helpful conversations.

\end{document}